\documentclass[oribibl]{llncs}

\PassOptionsToPackage{hyphens}{url}
\usepackage[colorlinks = true,
  linkcolor = blue,
  citecolor = blue,
  urlcolor = blue]{hyperref}

\usepackage{url}
\usepackage{csquotes}
\usepackage{multirow}
\usepackage{amsmath}
\usepackage{amssymb}
\usepackage{graphicx}
\usepackage{tabularx}
\usepackage{booktabs}
\usepackage{threeparttable}

\DeclareMathOperator{\Cor}{Cor}

\usepackage{fancyhdr} 
\fancypagestyle{firststyle}
{
\fancyhf{}
\fancyfoot[C]{\scriptsize{Proceedings of the 37th International Conference on Testing Software and Systems (ICTSS 2025), Limassol, Springer, pp.~325--332. This version is the authors' copy. The publisher's definite version is available online at Springer via \url{https://doi.org/10.1007/978-3-032-05188-2_21}.}}
}

\begin{document}

\title{Tracing Vulnerability Propagation \\ Across Open Source Software
  Ecosystems}

\author{Jukka Ruohonen\orcidID{0000-0001-5147-3084} \and Qusai Ramadan\orcidID{0000-0001-8159-918X} \\
  \email{\{juk, qura\}@mmmi.sdu.dk}}
\institute{University of Southern Denmark, S\o{}nderborg, Denmark}

\maketitle

\begin{abstract}
The paper presents a traceability analysis of how over 84 thousand
vulnerabilities have propagated across 28 open source software
ecosystems. According to the results, the propagation sequences have been
complex in general, although GitHub, Debian, and Ubuntu stand out. Furthermore,
the associated propagation delays have been lengthy, and these do not correlate
well with the number of ecosystems involved in the associated sequences. Nor
does the presence or absence of particularly ecosystems in the sequences yield
clear, interpretable patterns. With these results, the paper contributes to the
overlapping knowledge bases about software ecosystems, traceability, and
vulnerabilities.
\end{abstract}

\begin{keywords}
software ecosystems, vulnerabilities, traceability, process mining
\end{keywords}

\section{Introduction}

\thispagestyle{firststyle} 

The paper continues and advances the empirical vulnerability coordination
research theme~\text{\cite{Lin23, Ruohonen18IST}} in a software ecosystem
context; the interest is to better understand how vulnerabilities propagate
across ecosystems. The ecosystem context places the paper into a large branch
that has examined software and cyber security of particularly programming
language ecosystems, including with respect to a recent increase of malware
uploads to these ecosystems~\cite{Akhavani25, Ruohonen25ARES}. To narrow the
branch a little, the paper can be characterized to operate specifically in a
\textit{cross-ecosystem} context that has recently received further
attention~\cite{Wang24, Xu22}. In other words, the paper empirically examines
the propagation of vulnerabilities across multiple open source software
ecosystems. With respect to software traceability research, which too is a large
branch, the paper operates in a \text{\textit{post-release}} traceability
context because all software observed has already been released \cite{Dakkak22,
  Ruohonen17APSEC}. Patching and otherwise handling of vulnerabilities is a
classical example of typical post-release software engineering
activities~\cite{Dissanayake22, Zhang25}, and, hence, it is not a surprise that
also vulnerabilities have been traced~\cite{Alqahtani17, Alqahtani16,
  YangSmith25}.

The paper's practical relevance can be motivated by the Cyber Resilience
Act~(CRA) recently agreed upon in the European Union.\footnote{~Regulation (EU)
2024/2847.} Among the regulation's essential cyber security requirements are
legal obligations to follow a coordinated vulnerability disclosure policy,
supply security patches, and to only ship products without known
vulnerabilities. As open source software is widely used also in the commercial
software industry, fulfilling the obligations requires tracking, coordinating,
and tracing of vulnerabilities also with respect to open source software
components, including those distributed through software ecosystems. Finally and
regarding the remaining structure: Section~\ref{sec: data and methods} presents
the data and methods used, Section~\ref{sec: results} the results, and
Section~\ref{sec: conclusion} a conclusion and a discussion.

\section{Data and Methods}\label{sec: data and methods}

\subsection{Data}

The dataset examined is based on the new Open Source Vulnerabilities (OSV)
database.\footnote{~\url{https://osv.dev/}} It has been used in recent
research~\cite{Ruohonen25ARES, Zhang25}, and it is ideal for the paper's
purposes because it provides cross-ecosystem traceability data. The dataset was
limited to vulnerabilities (such that malware entries were omitted) archived
with~Common Vulnerabilities and Exposures (CVEs). Regarding the ecosystems, some
of which use custom identifiers mapped to CVEs, all unique CVEs were included
from the \texttt{aliases}, \texttt{upstream}, and \texttt{related} fields in the
OSV's JavaScript Object Notation (JSON) schema. Given data retrieval in 19 April
2025, the dataset contains $n = 84,520$ vulnerabilities and $28$ ecosystems.

A note should be also made about the ecosystem term; it refers to the
corresponding concept used in the OSV
database.\footnote{~\url{https://google.github.io/osv.dev/data/\#covered-ecosystems}}
These ecosystems cover programming language ecosystems, such as PyPI for Python
or Packagist for PHP, Linux distributions, such as Red Hat and Ubuntu, software
testing frameworks, such as OSS-Fuzz, and large hosting services, such as GitHub
for which GitHub Actions were merged with the main GitHub ecosystem
entries.\footnote{~Although OSV uses a term Git, the term GitHub is used for
clarity. However, it must be also acknowledged that the terminology used is a
limitation because it is not entirely clear to which the Git entries
specifically refer in all cases. This point serves also as a note that further
research is needed to validate the OSV database.} This terminological choice can
be justified on the grounds that there is no consensus in the literature about
definitions~\cite{Burnstrom24}. The choice also maintains coherence with OSV.

\subsection{Methods}

In overall, the methodology adopted is based on process
mining~\cite{Ghahderijani25, vanDerAalst22}. Accordingly, the propagation of
CVEs across software ecosystems are modeled as an event log. An event is
understood as a tuple:

\begin{equation}
e = (c, a, t) \in \mathcal{C} \times \mathcal{A} \times \mathbb{T} ,
\end{equation}
where $c \in \mathcal{C}$ is a unique CVE in a set of all CVEs,
$\vert\mathcal{C}\vert = n$, $a \in \mathcal{A}$ is a unique ecosystem in a set
of all ecosystems observed, $\vert\mathcal{A}\vert = 28$, and $\mathbb{T}
\subseteq \mathbb{N}$ is a set of discrete timestamps (measured in days). In
terms of years, the timestamps start from 2000 and end to mid-April 2025. Then,
an event log $L \subseteq \mathcal{C} \times \mathcal{A} \times \mathbb{T}$ is a
multiset of events. For any $c \in \mathcal{C}$, the trace for the given $c$ is
defined as:
\begin{equation}\label{eq: trace}
\sigma_c = \langle (a_1, t_1), (a_2, t_2), \ldots, (a_m, t_m) \rangle,
\quad \textmd{such that} \quad t_1 \leq t_2 \leq \ldots \leq t_m .
\end{equation}

The trace in \eqref{eq: trace} represents an ordered sequence through which a
given CVE appeared across unique ecosystems. If a length of a given sequence is
one, the given CVE-referenced vulnerability $c \in \mathcal{C}$ appeared only in
a single ecosystem.

Two assumptions are made:

\begin{enumerate}

\itemsep 3pt

\item{\textit{Independence of CVEs}: each CVE, $c \in \mathcal{C}$, is modeled as
  an independent case. There are no causal dependencies between different CVEs.
  Formally, for any two CVEs, $c_1, c_2 \in \mathcal{C}$ with $c_1 \neq c_2$, the
  traces $\sigma_{c_1}$ and $\sigma_{c_2}$ are assumed independent.
  Furthermore, the propagation of CVEs through ecosystems is treated as a
  Poisson process over discrete time, implying that time differences between
  events are realizations of a Poisson-distributed random variable.}

\item{\textit{Uniqueness of ecosystem observations}: for each $c \in
  \mathcal{C}$ and each ecosystem $a \in \mathcal{A}$, there exists at most one
  recorded event for the appearance of $c$ given~$a$.  In the presence of
  multiple observations, only the event with the earliest timestamp is
  considered.  Formally, for duplicated events $(c, a, t_i)$ and \( (c, a,
  t_j) \) with \( t_i \neq t_j \), only the event with \( \min(t_i, t_j) \) is
  retained.}

\end{enumerate}

The interest is to empirically observe partial ordering over ecosystems based on timestamp comparisons with a trace:
\begin{equation}\label{eq: propagation}
a_i \preceq a_j \iff t_i \leq t_j \quad\textmd{and}\quad \vert \sigma_c \vert > 1.
\end{equation}
This ordering allows observing symbolic representations of propagation
sequences across ecosystems. For instance:
\begin{equation}\label{eq: example sequence}
\textmd{GitHub} \preceq \textmd{Debian} \preceq \textmd{Ubuntu} ,
\end{equation}
which means that a given CVE was observed in GitHub first, then in Debian, and
then in Ubuntu, respecting the timestamps observed.

Already observing sequences such as \eqref{eq: example sequence} visually with
descriptive statistics provides valuable insights into the propagation of
CVE-referenced vulnerabilities across the twenty-eight ecosystems tracked in the
OSV database. In addition, the interest is to observe traceability delays,
as defined by:

\begin{equation}\label{eq: delays}
f(c) = t_{\text{last}} - t_{\text{first}} ,
\end{equation}
where $t_{\text{first}}$ and $t_{\text{last}}$ are the first and last timestamps
in a trace $\sigma_c$ longer than $\vert \sigma_c \vert > 1$. The function
measures how long it took for a $c \in \mathcal{C}$ to spread from its first
known appearance to its last known appearance across all ecosystems.

The function also allows to formalize two \textit{hypotheses}
($\textmd{H}$). First ($\textmd{H}_1$), it seems sensible to assume that
$\Cor(f(c), \vert \sigma_c \vert) > 0$ for the $m = \vert \sigma_c \vert > 1$
sequences, meaning that the more there are ecosystems, the longer the
traceability delays. Second~($\textmd{H}_2$), it can be hypothesized that $f(c)$
might vary according to some particular ecosystems that appear in the $\vert
\sigma_c \vert > 1$ sequences. For instance, the presence of some particular
ecosystem might shorten the traceability delays, whereas the appearance of some
other ecosystem might indicate a bottleneck in terms of delays. To examine this
$\textmd{H}_2$, for each $a \in \mathcal{A}$, a $t$-test (with a correction for
unequal variances; \cite{Welch47}) is computed for testing whether the presence
of a given $a$ in a $\sigma_c$ affects the corresponding $f(c)$. Due to repeated
testing, Bonferroni correction is applied, meaning that the statistical
significance level is set to $0.05~/~\vert\mathcal{A}\vert$. Although much
debated, the Bonferroni correction seems suitable because many tests are carried
out, $\textmd{H}_2$ is fairly loose in theoretical terms, and the purpose is to
check whether any ecosystem shows a statistically significant
relationship~\cite{Amstrong14}. With these elaborations, the results can be
disseminated next.

\section{Results}\label{sec: results}

\begin{figure}[th!b]
\centering
\includegraphics[width=\linewidth, height=10.5cm]{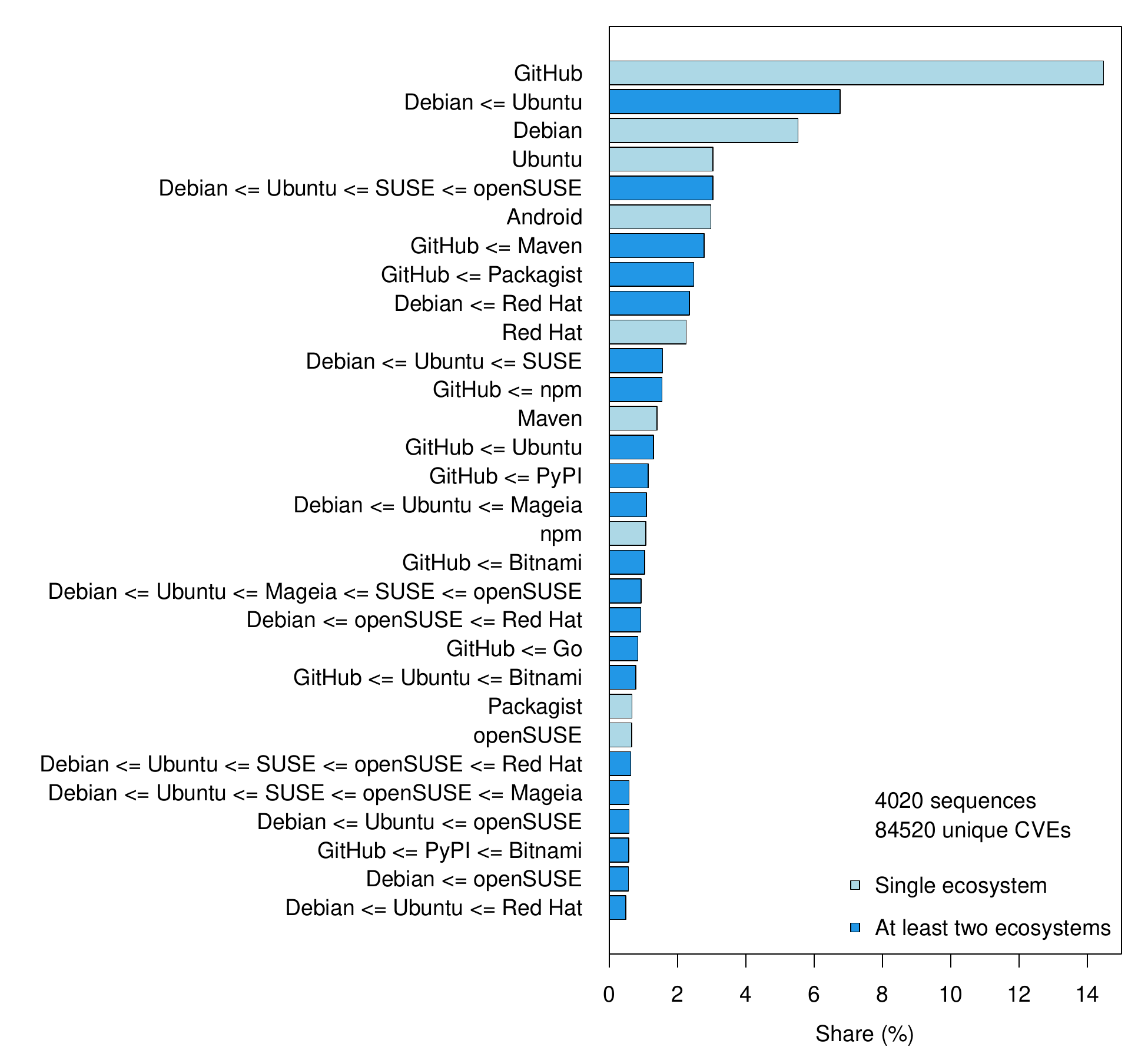}
\caption{The Top-30 Sequences Extracted}
\label{fig: top sequences}
\end{figure}

The sequences elaborated in \eqref{eq: propagation} and \eqref{eq: example
  sequence} provide a good way to start the dissemination of the empirical
results. Thus, the top-30 sequences (in terms of frequency) are shown in
Fig.~\ref{fig: top sequences}. (For presentation purposes, the figure shows also
cases with $\vert\sigma_c\vert = 1$.) In total, a little over four thousand
unique sequences were identified for the about $85$ thousand unique CVEs. This
amount alone testifies that the propagation can be quite complex among the $28$
ecosystems sampled. Both the mean and median are six ecosystems per a CVE on
average. However, GitHub alone leads the ranking in Fig.~\ref{fig: top
  sequences}; nearly $14\%$ of the CVEs observed were reported only in GitHub
and nowhere else. GitHub is also present in many traces with a
$\vert\sigma_c\vert = 2$, meaning that many other ecosystems, including both
programming language ecosystems and Linux distributions, have tended to pick
CVE-referenced vulnerabilities that were initially reported on GitHub.

\begin{figure}[th!b]
\centering
\includegraphics[width=\linewidth, height=4.5cm]{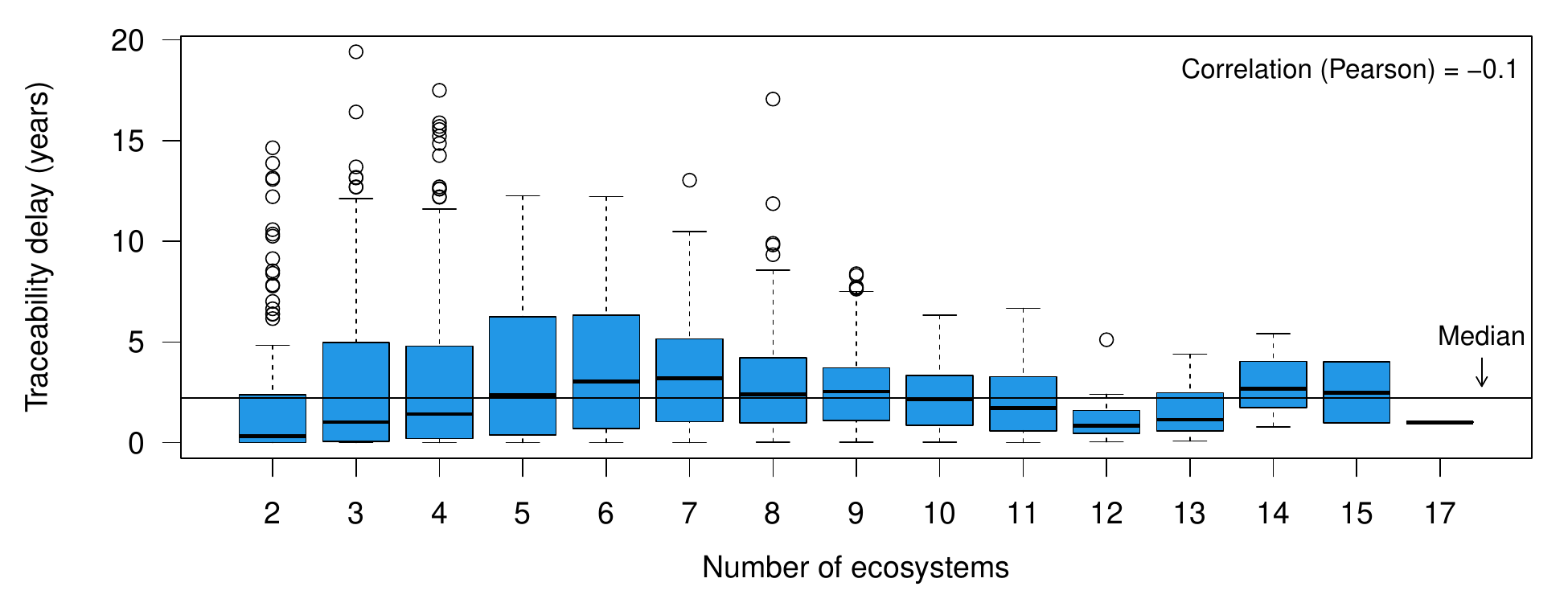}
\caption{Traceability Delays ($55,845$ unique CVEs for the $3,997$ sequences with $\vert\sigma_c\vert > 1$)}
\label{fig: diff}
\end{figure}

The second place in the ranking is taken by CVEs first reported in Debian and
then in Ubuntu. This observation is logical because most of Ubuntu's packages
are distributed also in Debian, and largely also maintained therein. Having said
that, there are also quite a few CVEs that were reported either in Debian or
Ubuntu alone. Although further validation would be required, this observation
could be taken to support an argument that there might be still room for better
coordination and synchronization. The same point extends toward other Linux
distributions who too seem to often pick CVEs that were first reported
in~Debian. In this regard, it also worth emphasizing that Debian, Ubuntu, Red
Hat, and SUSE, among a few others, such as the Python Software Foundation who
maintains PyPI, are official CVE numbering authorities (CNAs) unlike Alpine
Linux, Mageia, Rocky Linux, and some
others.\footnote{~\url{https://www.cve.org/ProgramOrganization/CNAs}} Because
the dataset's scope was restricted to CVE-referenced vulnerabilities, and
because CVEs require coordination of their own~\cite{Ruohonen18IST}, the
propagation sequences observed may be partially---but unlikely fully---explained
by the CNAs present in the sequences.

Turning to $\textmd{H}_1$, Fig.~\ref{fig: diff} shows the traceability delays,
as given by \eqref{eq: delays} for all sequences with lengths larger than one,
against the number of ecosystems, $m$, present in these sequences. The
hypothesis is rejected; the Pearson's product moment correlation coefficient is
small in magnitude and has a negative sign. In other words---and unlike what was
expected and what would seem intuitively logical, the traceability delays do not
notably shorten or lengthen according to whether there are many or a few
ecosystems present in the traces. Relatively long propagation sequences, such as
those with five Linux distributions in Fig.~\ref{fig: top sequences}, are not
necessarily slower than shorter ones. In addition to this observation,
Fig.~\ref{fig: diff} delivers an important point: the traceability delays have
generally been long on average. The mean and median are as long as three and two
years, respectively. As can be further seen from the figure, there are also a
lot of outliers, including even extreme ones indicating delays over a
decade. Although the reasons for such outliers are not well-known, similar
observations have been made previously~\cite{Ruohonen18IST}. Regardless of the
potential explanations, the observation reinforces the earlier remark about
potential gains from better coordination and synchronization.

Regarding $\textmd{H}_2$, the Bonferroni-corrected and variance-adjusted
$t$-tests indicate that for ten ecosystems out of the differences between the
means are not different from zero, meaning that a presence or an absence of a
given $a$ does not affect the averages of the given traceability
delays.\footnote{~CRAN was excluded because only a single CVE has been reported
for it.} Among these ten ecosystems are GitHub, Maven, openSUSE, Packagist,
PyPI, RubyGems, and Ubuntu. In contrast, statistically significant differences
are present for Debian, Mageia, Red Hat, and SUSE for which their presences
indicate longer delays. If the propagation sequences have involved Go or npm,
the delays have been shorter. In general, these test results indicate that there
is no clear-cut pattern among the ecosystems with respect to the traceability
delays, including regarding the CNAs.

\section{Conclusion}\label{sec: conclusion}

This short paper presented an empirical traceability analysis of the propagation
of CVEs across popular open source software ecosystems, including Linux
distributions and programming language ecosystems. The analysis presented
demonstrates the value offered by the new OSV database also for research
purposes.

Regarding the results, (1)~the propagation has generally been rather complex, as
demonstrated by over four thousand unique propagation sequences for the $28$
ecosystems and about $85$ thousand unique CVEs observed. In terms of a frequency
ranking, however, (2)~GitHub alone, Debian and Ubuntu together, and Debian and
Ubuntu alone lead the ranking, meaning that also many CVEs have only been
reported in one ecosystem without a propagation to others. The frequent
propagation from Debian to Ubuntu is also expected, given the close resemblance
and collaboration between these two Linux distributions. Furthermore, (3)~the
traceability delays, as measured by time differences between the last and first
appearances of CVEs in given ecosystems, do not correlate well with the number
of ecosystems present in the corresponding propagation sequences. That is,
longer (shorter) sequences do not imply lengthier (faster) delays. Although
reporting was omitted for brevity, it can be noted that neither the severity of
the CVE-referenced vulnerabilities, as measured by the Common Vulnerability
Scoring System (v.~3.1), affects the traceability delays
statistically.\footnote{~\url{https://www.first.org/cvss/v3-1/}} In addition,
(4) there is no clear---or at least easily interpretable---pattern in the
traceability delays with respect to some particular ecosystems appearing in the
sequences. Last but not least, (5) the traceability delays have been lengthy on
average; the median is about two years. There are also extremely outlying
traceability delays.

Regarding future research, a good and relevant research topic would involving
validating the OSV database. In other words, (a)~it remains unclear how accurate
and generally robust the database is. This uncertainty affects also the results
reported. In addition, (b)~a notable limitation of the paper is that not all
fields were included in the data extraction. The \texttt{affected} field is
worth mentioning explicitly in this regard. Likewise, (c)~the focus on
CVE-referenced vulnerabilities can be mentioned as a limitation, although it
remains unclear whether and how cross-ecosystem empirical analysis could be
pursued without unique identifiers. Given the recent surge of malware uploads to
some of the ecosystems, as also archived in the OSV
database~\cite{Ruohonen25ARES}, (d)~it might make also sense to extend an
empirical analysis beyond vulnerabilities. In a similar vein, (e)~it would seem
reasonable to extend the propagation concept. In this regard, the OSV database
archives also security advisories with the \texttt{references} field. Taking
these and possibly other traces into account would perhaps give a more complete
picture of the propagation dynamics across ecosystems and open source software
projects.

Addressing these topics would help at eventually tackling the enduring
challenges related to data quality and vulnerability provenance
meta-data~\cite{Bohme25}. The OSV database has already facilitated a better
collection of meta-data but further work is required to put the meta-data
collected into use. By retrieving and validating data referenced in the OSV's
traceability links, it would possible to continue toward a more elaborate
propagation analysis.  In light of the rather long propagation delays, it could
be perhaps even contemplated whether the OSV database could (or should) be
developed further into a more general platform for coordinating and otherwise
handling of vulnerabilities in the open source software
context~\cite{Zhang25}. Although much of development nowadays happens on GitHub,
it remains unclear how well the platform works in facilitating coordination,
especially when keeping in mind the large amounts of software distributed
through the twenty-eight open source software ecosystems covered in the paper.

\bibliographystyle{splncs03}

\end{document}